\documentclass[a4paper,10pt]{article}
\usepackage{amssymb}
\usepackage{amsmath}
\usepackage{epsfig}
\usepackage{subfigure}
\usepackage{graphics}

\usepackage{latexsym}
\usepackage{rotating}
\usepackage{titlesec}

\title{Discrete Symmetries and Nonlocal Reductions}

\author{Metin G{\" u}rses \thanks{
email: gurses@fen.bilkent.edu.tr} \\
{\small Department of Mathematics, Faculty of Sciences}\\
 {\small Bilkent University, 06800 Ankara, Turkey}~~~~\\
 Asl{\i} Pekcan \thanks{
email: aslipekcan@hacettepe.edu.tr}\\
{\small Department of Mathematics},\\
{\small Hacettepe University, 06800 Ankara, Turkey}~~~~\\
 Konstyantyn Zheltukhin \thanks{
email: zheltukh@metu.edu.tr}\\
{\small Department of Mathematics},\\
{\small Middle East Technical University, 06800 Ankara, Turkey}}

\date{\nonumber}
\begin{document}
\maketitle
\date{\nonumber}
\newtheorem{thm}{Theorem}[section]
\newtheorem{Le}{Lemma}[section]
\newtheorem{defi}{Definition}[section]
\newtheorem{ex}{Example}[section]
\newtheorem{pro}{Proposition}[section]
\baselineskip 17pt

\numberwithin{equation}{section}

\begin{abstract}
We show that nonlocal reductions of systems of integrable nonlinear partial differential equations are the special discrete symmetry transformations.\\

\end{abstract}

\section{Introduction}

Nonlocal reductions of systems of integrable nonlinear partial differential equations which were invented first by Ablowitz and Musslimani \cite{AbMu1}-\cite{AbMu3}, attracted many researchers in the field. Ablowitz and Musslimani have first constructed nonlocal reduction for nonlinear Schr\"{o}dinger (NLS) system of equations and obtained nonlocal nonlinear Schr\"{o}dinger (nNLS) equation \cite{AbMu1}, \cite{AbMu2}. They showed that nNLS equation is integrable, i.e.,  it admits a Lax pair, and found soliton solutions by the use of the inverse scattering method. Ablowitz and Musslimani have later extended their nonlocal reductions, corresponding to space reflection, time reflection, and space-time reflection to modified Korteweg-de Vries (mKdV) system, sine-Gordon (SG) system, Davey-Stewartson (DS) system and so on. After Ablowitz and Musslimani's works there is a huge interest in obtaining  nonlocal reductions of systems of integrable equations and finding  interesting wave solutions of these systems. Specific examples are nonlocal NLS equation \cite{AbMu1}-\cite{jianke}, nonlocal  mKdV equation \cite{AbMu2}-\cite{chen}, \cite{GurPek3}, \cite{GurPek2}-\cite{ma}, nonlocal SG equation \cite{AbMu2}-\cite{chen}, \cite{aflm}, nonlocal DS equation \cite{AbMu3}, \cite{fok}-\cite{ZL}, nonlocal Fordy-Kulish equations \cite{GurPek3}, \cite{GursesFK}, nonlocal $N$-wave systems \cite{AbMu3}, \cite{gerd2}, nonlocal vector NLS equations \cite{sin}-\cite{gerd3}, nonlocal $(2+1)$-dimensional negative AKNS systems \cite{GurPek4}, nonlocal coupled Hirota-Iwao mKdV systems \cite{Pek}. See \cite{super} for the discussion of superposition of nonlocal integrable equations, and \cite{hydro} for the nonlocal reductions of the integrable equations of hydrodynamic type.
The connection between local and nonlocal reductions is given in \cite{Vincent}, \cite{Yang}.
In all these works the soliton solutions and their properties were investigated by using the inverse scattering method, by the Hirota bilinear method, and by Darboux transformations.

In the last decade we observe that even as the number of systems of integrable nonlinear differential equations possessing nonlocal reductions is increasing, there is no one so far explaining how or where such nonlocal reductions come from.  The origin of nonlocal reductions was mysterious.
In this work we address to this problem. We show that those systems possessing nonlocal reductions admit discrete symmetry transformations which leave the systems invariant. A special case of discrete symmetry transformation turns out to be the nonlocal reductions of the same systems. We show this fact for NLS, mKdV, SG, DS, coupled NLS-derivative NLS, loop soliton systems, hydrodynamic type systems, and Fordy-Kulish equations, and derive all possible nonlocal reductions from the discrete symmetry transformations of these systems.

\section{Reductions}
\noindent Let the dynamical variables $q^{i}(t,x)$ and $r^{i}(t,x)$  ($i=1,2,\cdots, N$), in $(1+1)$-dimensions, satisfy the following system
of integrable evolution equations
\begin{eqnarray}
q^{i}_{t}&=&F^{i}(q^{j}, r^{j}, q^{j}_{x}, r^{j}_{x}, q^{j}_{xx}, r^{j}_{xx}, \cdots),~~~ i, j=1, 2, \cdots, N , \label{cdenk1}\\
 \nonumber \\
r^{i}_{t}&=&G^{i}(q^{j}, r^{j}, q^{j}_{x}, r^{j}_{x}, q^{j}_{xx}, r^{j}_{xx}, \cdots),~~~ i, j=1, 2, \cdots, N,  \label{cdenk2}
\end{eqnarray}
where $F^{i}$ and $G^{i} ~(i=1,2,\cdots,N)$ are functions of the dynamical variables $q^{i}(t,x)$, $r^{i}(t,x)$, and their partial derivatives with respect to $x$. The above system of equations is integrable, so it has a Lax pair and a recursion operator $\mathcal{R}$.
Some of these equations admit local and nonlocal reductions. Let us assume that the above system of equations (\ref{cdenk1}) and (\ref{cdenk2}) admits the following reductions.

\vspace{0.3cm}
\noindent
(a) Local reductions:

\noindent The local reductions are given by
\begin{equation}\label{generalred1}
r^{i}(t,x)=\kappa_{1}\, q^{i}(t,x),\quad i=1, 2, \cdots, N,
\end{equation}
and
\begin{equation}\label{generalred2}
r^{i}(t,x)=\kappa_{2}\, \bar{q}^{i}(t,x),\quad i=1, 2, \cdots, N,
\end{equation}
where $\kappa_{1}$ and $\kappa_{2}$ are real constants. Throughout this paper a bar over a letter is defined as
\begin{enumerate}
\item[1)] for a complex number $q=\alpha+i\beta$, $\bar{q}=\alpha-i\beta$, $i^2=-1$,
\item[2)] for a pseudo-complex number $q=\alpha+i\beta$,  $\bar{q}=\alpha-i\beta$, $i^2=1$.
 \end{enumerate}
If a reduction is consistent the system of equations (\ref{cdenk1}) and (\ref{cdenk2}) is reduced to a system for $q^{i}$'s
\begin{equation}\label{first}
q^{i}_{t}={\tilde F}^{i}(q^{j}, q^{j}_{x}, q^{j}_{xx}, \cdots), \quad i, j=1,2,\cdots, N
\end{equation}
for the reduction (\ref{generalred1}) and
\begin{equation}\label{second}
q^{i}_{t}={\tilde F}^{i}(q^{j}, \bar{q}^{j}, q^{j}_{x}, \bar{q}_x^{j}, q^{j}_{xx},\bar{q}_{xx}^j, \cdots), \quad i, j=1,2,\cdots, N
\end{equation}
for the reduction (\ref{generalred2}), where ${\tilde F}=F|_{r^{i}=\kappa_1 q^{i}}$ or ${\tilde F}=F|_{r^{i}=\kappa_2 \bar{q}^{i}}$, respectively.

\vspace{0.3cm}
\noindent
{\bf (b)} Nonlocal reductions:

\noindent Recently, Ablowitz and Musslimani introduced new type of reductions \cite{AbMu1}-\cite{AbMu3}
\begin{equation}
r^{i}(t,x)=\tau_{1} q^{i}(\varepsilon_{1} t, \varepsilon_{2} x)=\tau_1q_{\varepsilon}^i, \label{red1}
\end{equation}
and
\begin{equation}
r^{i}(t,x)=\tau_{2} \bar{q}^{i}(\varepsilon_{1} t, \varepsilon_{2} x)=\tau_2\bar{q}_{\varepsilon}^i, \label{red2}
\end{equation}
for $i=1,2,\cdots,N$. Here $\tau_{1}$ and $\tau_{2}$  are real constants and $\varepsilon_{1}^2=\varepsilon_{2}^2=1$.

\noindent When
$(\varepsilon_{1}, \varepsilon_{2})=(-1,1),(1,-1),(-1,-1)$ the above constraints reduce the system (\ref{cdenk1}) and (\ref{cdenk2}) to nonlocal space reflection symmetric (S-symmetric), time reflection symmetric (T-symmetric), or space-time reflection symmetric (ST-symmetric) differential equations.

\noindent
Since the reductions are done consistently  the reduced systems of equations are  also  integrable. This means that the reduced systems
admit recursion operators and Lax pairs.  We can obtain $N$-soliton solutions of the reduced systems by the inverse scattering method  \cite{AbMu1}-\cite{AbMu3}, \cite{aflm1}, \cite{Wen}, \cite{jianke}, \cite{JZ2}, \cite{aflm}, \cite{sin}, by the Darboux transformation \cite{li}, \cite{JZ1}, \cite{ma}, \cite{XLHCY}, \cite{ZXZhou}, and by the Hirota bilinear method \cite{GurPek1}, \cite{GurPek3}, \cite{GurPek2}, \cite{RZFH}, \cite{GurPek4}-\cite{super}.

\section{Discrete Symmetries}

In this section we will show that nonlocal reductions arise from scaling symmetries of integrable system  of equations. A scaling symmetry of a system of differential equations is the scale transformation which leaves these equations invariant. Scaling symmetries group is  a subgroup of the symmetry groups  of differential equations \cite{olv} and discrete symmetries are special cases of the scaling symmetries \cite{hydon}.\\

\noindent \textbf{(a)}\, \textbf{NLS System}: This system is given by
\begin{eqnarray}
&& a q_{t}= -\frac{1}{2}\, q_{xx}+q ^2\, r,\label{eq3} \\
&& a r_{t}=\frac{1}{2}\, r_{xx}-q \, r^2, \label{eq4}
\end{eqnarray}
where $a$ is any constant. This constant is the imaginary unit for the original NLS system but we change it by redefining the $t$ variable.
We search for a symmetry transformations such that the NLS system is left invariant. In general  we choose the symmetry transformation  as
$$T_{1}: (q(t,x),r(t,x)) \to (q^{\prime}(t^{\prime}, x^{\prime}), r^{\prime}(t^{\prime}, x^{\prime}))$$
where primed system satisfies also the NLS system, i.e.,
\begin{eqnarray}
&& a q^{\prime}_{t^{\prime}}= -\frac{1}{2}\, q^{\prime}_{x^{\prime} x^{\prime}}+(q^{\prime}) ^2\, r^{\prime}, \label{denk3} \\
&& a r^{\prime}_{t^{\prime}}=\frac{1}{2}\, r^{\prime}_{x^{\prime} x^{\prime}}-q^{\prime} \, (r^{\prime})^2. \label{denk4}
\end{eqnarray}
We shall consider the real and complex dynamical systems separately. For the real case
the symmetry transformation that we are interested in is the scale transformations
\begin{eqnarray}
&&t^{\prime}=\beta\,t,~~~x^{\prime}=\alpha \,x, \\
&&q^{\prime}=\gamma_{1}\,q+\delta_{1}\,r,\\
&&r^{\prime}=\gamma_{2}\, r+\delta_{2}\,q,
\end{eqnarray}
where $\alpha, \beta, \gamma_{1}, \gamma_{2}, \delta_{1}$, and $\delta_{2}$ are real constants.
We have two possible cases:

\vspace{0.3cm}
\noindent
(a)\, First type of real scale symmetry transformation is
\begin{eqnarray}
&&t^{\prime}=-\alpha^2\,t,~~~x^{\prime}=\alpha \,x, \\
&&q^{\prime}=\delta_{1}\,r,\\
&&r^{\prime}=\frac{1}{\delta_{1}\, \alpha^2}\,q,
\end{eqnarray}
where $\alpha$ and $\delta_{1}$ are arbitrary constants.

\vspace{0.3cm}
\noindent
(b)\, Second type of real scale symmetry transformation is
\begin{eqnarray}
&&t^{\prime}=\alpha^2\,t,~~~x^{\prime}=\alpha \,x, \\
&&q^{\prime}=\gamma_{1}\,q,\\
&&r^{\prime}=\frac{1}{\gamma_{1}\, \alpha^2}\,r,
\end{eqnarray}
where $\alpha$ and $\gamma_{1}$ are arbitrary constants.
These two parameter transformations map solutions to solutions of the NLS system.

\vspace{0.3cm}
\noindent
From the above scale symmetry transformation we can obtain discrete symmetry transformations by letting $\alpha=\epsilon=\pm1$.
In particular the first type produces a discrete symmetry transformation if $\alpha=\epsilon$ and $\delta_{1}=k$ then
\begin{eqnarray}
&&q(t,x)=k\, r^{\prime}(-t, \epsilon x), \\
&&r(t,x)=k\, q^{\prime}(-t, \epsilon x),
\end{eqnarray}
where $\epsilon^2=k^2=1$. A special discrete symmetry transformation is obtained
when we take $q^{\prime}=q$ and $r^{\prime}=r$.
 This special discrete symmetry is the well-known nonlocal reductions $\displaystyle r(t,x)=k q(-t,x)$ and $\displaystyle r(t,x)=k q(-t,-x)$ \cite{AbMu3}, \cite{chen}, \cite{gerd}, \cite{aflm1}, \cite{GurPek3}, \cite{jianke}.

\noindent For the complex dynamical systems the scale symmetry transformation
$$T_{2}:(\bar{q}(t,x),\bar{r}(t,x))\to (q^{\prime}(t^{\prime}, x^{\prime}), r^{\prime}(t^{\prime}, x^{\prime}))$$
 takes the following form
\begin{eqnarray}
&&t^{\prime}=\beta\,t,~~~x^{\prime}=\alpha \,x, \\
&&q^{\prime}=\gamma_{1}\,\bar{q}+\delta_{1}\,\bar{r},\\
&&r^{\prime}=\gamma_{2}\, \bar{r}+\delta_{2}\,\bar{q}.
\end{eqnarray}
where $\alpha, \beta, \gamma_{1}, \gamma_{2}, \delta_{1}$ and $\delta_{2}$ are real constants.
We have two possible cases:

\vspace{0.3cm}
\noindent
(a)\, First type of complex scale symmetry transformation is
\begin{eqnarray}
&&t^{\prime}=\beta\,t,~~~x^{\prime}=\alpha \,x, \\
&&q^{\prime}=\delta_{1}\,\bar{r},\\
&&r^{\prime}= \delta_{2}\,\bar{q},
\end{eqnarray}
with
\begin{equation}\label{const1}
\bar{a}\, \beta=-a \alpha^2,~~\delta_{1}\, \delta_{2} \, \alpha^2=1.
\end{equation}

\vspace{0.3cm}
\noindent
(b)\, Second type of complex scale symmetry transformation is
\begin{eqnarray}
&&t^{\prime}=\beta \,t,~~~x^{\prime}=\alpha \,x, \\
&&q^{\prime}=\gamma_{1}\,\bar{q},\\
&&r^{\prime}=\gamma_{2}\,\bar{r},
\end{eqnarray}
with
\begin{equation}
\bar{a}\, \beta=a \alpha^2,~~ \gamma_{1}\, \gamma_{2} \, \alpha^2=1.\,
\end{equation}
These two parameter transformations map also solutions to solutions of the NLS system. From these scale symmetry transformations we obtain
discrete symmetry transformation by letting $\alpha=\epsilon_{1}=\pm 1$, $\beta=\epsilon_{2}=\pm 1$, $\gamma_{1}=\gamma_{2}=k=\pm 1$.
In particular the first type produces a discrete symmetry transformation of the form
\begin{eqnarray}
&&q(t,x)=k\, \bar{r}^{\prime}(\epsilon_{2} t, \epsilon_{1}  x), \\
&&r(t,x)=k\, \bar{q}^{\prime}( \epsilon_{2} t, \epsilon_{1} x),
\end{eqnarray}
where $\epsilon_{1}^2=\epsilon_{2}^2=k^2=1$ and $\bar{a} \epsilon_{2}=-a$ which follows from (\ref{const1}). A special discrete symmetry transformation is obtained when
we take $q^{\prime}=q$ and $r^{\prime}=r$. This special symmetry  is the well-known  nonlocal reductions $\displaystyle r(t,x)=k\, \bar{q}( -t, x)$ with $\bar{a}=-a$,  $\displaystyle r(t,x)=k\, \bar{q}( t, -x)$ with $\bar{a}=a$, and $\displaystyle r(t,x)=k\, \bar{q}( -t, -x)$ with $\bar{a}=-a$ \cite{AbMu1}, \cite{AbMu2}, \cite{chen}-\cite{li}, \cite{Wen}-\cite{jianke}.

\vspace{0.3cm}
\noindent
The examples that we consider in the rest of the paper share similar real and complex scale symmetry transformations and the associated discrete symmetry transformations. Since we are interested in nonlocal reductions of the integrable systems of equations we will present only the first type real and complex discrete transformations and the corresponding nonlocal reductions.

\vspace{0.3cm}
\noindent \textbf{(b)}\, \textbf{MKdV System}:  This system is given by
\begin{eqnarray}
&& a q_{t}= -\frac{1}{4}\, q_{xxx}+\frac{3}{2} q \, r\, q_{x}, \label{eq16} \\
&& a r_{t}=-\frac{1}{4}\, r_{xxx}+\frac{3}{2} q \, r \,r_{x}. \label{eq17}
\end{eqnarray}
 We will write the discrete symmetry transformations directly. We have two different cases: Let $(q,r)$ and $(q^{\prime}, r^{\prime})$ satisfy the mKdV system of equations (\ref{eq16}) and (\ref{eq17}).

\vspace{0.3cm}
\noindent
For the real case we have
\begin{equation}
q(t,x)=k r^{\prime}( \epsilon_{2} t, \epsilon_{1} x), ~~ r(t,x)=k q^{\prime}( \epsilon_{2} t, \epsilon_{1} x), \label{eq18}
\end{equation}
where $k^2=1$ and $\epsilon_{1} \epsilon_{2}=1$. When we take $q^{\prime}=q$ and $r^{\prime}=r$ we obtain the nonlocal reduction $\displaystyle r(t,x)=k\, q( -t, -x)$ \cite{AbMu2}-\cite{chen}, \cite{GurPek3}, \cite{GurPek2}-\cite{JZ2}.

\vspace{0.3cm}
\noindent
For  the complex case we have
\begin{equation}
q(t,x)=k \bar{r}^{\prime}( \epsilon_{2} t, \epsilon_{1} x), ~~ r(t,x)=k \bar{q}^{\prime}( \epsilon_{2} t, \epsilon_{1} x), \label{eq19}
\end{equation}
where $\bar{a}\,\epsilon_{1} \epsilon_{2}=a$ and $k^2=1$. These special discrete transformations produce different nonlocal reductions when $q^{\prime}=q$ and $r^{\prime}=r$ with different values of $\epsilon_{1}=\pm 1$ and $\epsilon_{2}=\pm 1$; $\displaystyle r(t,x)=k\, \bar{q}( -t, x)$ with $\bar{a}=-a$,  $\displaystyle r(t,x)=k\, \bar{q}( t, -x)$ with $\bar{a}=-a$, and $\displaystyle r(t,x)=k\, \bar{q}( -t, -x)$ with $\bar{a}=a$ \cite{AbMu2}-\cite{chen}, \cite{GurPek3}, \cite{GurPek2}, \cite{ma}.

\vspace{0.3cm}

\noindent \textbf{(c)}\, \textbf{SG System}: This system is given by
\begin{eqnarray}
&& q_{xt}+2s\,q=0, \label{eq20} \\
&& r_{xt}+2s\,r=0, \label{eq21}\\
&& s_x+(q\,r)_t=0, \label{eq22}
\end{eqnarray}
where $q=q(t,x)$, $r=r(t,x)$, and $s=s(t,x)$. We have the following two discrete symmetry transformations.
For the real case,
\begin{equation}
q(t,x)=k r^{\prime}( \epsilon_{2} t, \epsilon_{1} x),\quad r(t,x)=k q^{\prime}( \epsilon_{2} t, \epsilon_{1} x),
\quad s(t,x)=s^{\prime}( \epsilon_{2} t, \epsilon_{1} x),
\end{equation}
where $\epsilon_{1}=\epsilon_2=\pm 1$ and $k^2=1$. If we take $q^{\prime}=q$ and $r^{\prime}=r$ these special discrete transformations produce the nonlocal reductions:
$r(t,x)=k q(-t,x)$,  $r(t,x)=k q(t,-x)$, and $r(t,x)=k q(-t,-x)$ \cite{AbMu2}-\cite{chen}, \cite{aflm}.

\noindent For the complex case,
\begin{equation}
q(t,x)=k \bar{r}^{\prime}( \epsilon_{2} t, \epsilon_{1} x), \quad  r(t,x)=k \bar{q}^{\prime}( \epsilon_{2} t, \epsilon_{1} x), \quad s(t,x)=\bar{s}^{\prime}( \epsilon_{2} t, \epsilon_{1} x),
\end{equation}
where $\epsilon_{1}=\epsilon_2=\pm 1$ and $k^2=1$. When $q^{\prime}=q$ and $r^{\prime}=r$ these special discrete transformations produce the nonlocal reductions:
$r(t,x)=k \bar{q}(-t,x)$,  $r(t,x)=k \bar{q}(t,-x)$, and $r(t,x)=k \bar{q}(-t,-x)$ \cite{chen}.

\vspace{0.3cm}

\noindent \textbf{(d)}\, \textbf{DS System}: This system is given by
\begin{eqnarray}
aq_t+\frac{1}{2}[\gamma^2q_{xx}+q_{yy}]+q^2r&=&\phi q, \\
-ar_t+\frac{1}{2}[\gamma^2r_{xx}+r_{yy}]+r^2q&=&\phi r, \\\
\phi_{xx}-\gamma^2\phi_{yy}&=&2(qr)_{xx},
\end{eqnarray}
where $q=q(t,x,y)$, $r=r(t,x,y)$, $\phi=\phi(t,x,y)$, $\gamma^2=\pm 1$, and $a$ is a constant. We have the following discrete symmetry transformations.
For the real case,
\begin{eqnarray}
&&q(t,x,y)=k r^{\prime}( \epsilon_{1} t, \epsilon_{2} x, \epsilon_{3} y), \\
&&r(t,x,y)=k q^{\prime}( \epsilon_{1} t, \epsilon_{2} x, \epsilon_{3} y), \\
&&\phi(t,x,y)=\phi^{\prime}( \epsilon_{1} t, \epsilon_{2} x, \epsilon_{3} y),
\end{eqnarray}
where $\epsilon_1=-1$ and $k^2=1$. These special discrete transformations produce the nonlocal reductions when $q^{\prime}=q$ and $r^{\prime}=r$ with different values of $\epsilon_{1}=- 1$, $\epsilon_{2}=\pm 1$, $\epsilon_{3}=\pm 1$; $\displaystyle r(t,x,y)=k\, q( -t, x, y)$,  $\displaystyle r(t,x,y)=k\, q( -t, -x,y)$, $\displaystyle r(t,x,y)=k\, q( -t, x,-y)$, and $\displaystyle r(t,x,y)=k\, q( -t, -x, -y)$ \cite{AbMu3}.

\noindent For the complex case,
\begin{eqnarray}
&&q(t,x,y)=k \bar{r}^{\prime}( \epsilon_{1} t, \epsilon_{2} x, \epsilon_{3} y), \\
&&r(t,x,y)=k \bar{q}^{\prime}( \epsilon_{1} t, \epsilon_{2} x, \epsilon_{3} y), \\
&&\phi(t,x,y)=\bar{\phi}^{\prime}( \epsilon_{1} t, \epsilon_{2} x, \epsilon_{3} y),
\end{eqnarray}
where $k^2=1$ , $\epsilon_{1}^2=\epsilon_{2}^2=\epsilon_{3}^2=1$, and $\bar{a} \epsilon_{1}=-a$. We observe that these discrete transformations produce many different nonlocal reductions when $q^{\prime}=q$, $r^{\prime}=r$, and $\phi^{\prime}=\phi$ with different values of $\epsilon_{1}=\pm 1$, $\epsilon_{2}=\pm 1$, and $\epsilon_{3}=\pm 1$; $\displaystyle r(t,x,y)=k\, \bar{q}( -t, x, y)$, $\displaystyle r(t,x,y)=k\, \bar{q}( -t, -x, y)$, $\displaystyle r(t,x,y)=k\, \bar{q}( -t, x, -y)$, $\displaystyle r(t,x,y)=k\, \bar{q}( -t, -x, -y)$   with $\bar{a}=a$;  $\displaystyle r(t,x,y)=k\, \bar{q}( t, -x, y)$, $\displaystyle r(t,x,y)=k\, \bar{q}( t, x, -y)$, $\displaystyle r(t,x,y)=k\, \bar{q}( t, -x, -y)$ with $\bar{a}=-a$ \cite{AbMu3}, \cite{fok}-\cite{ZL}.

\vspace{0.3cm}
\noindent \textbf{(e)}\, \textbf{Coupled NLS-derivative NLS System}: This system \cite{AbCl} is given by
\begin{eqnarray}
&&aq_t=iq_{xx}+\alpha(rq^2)_x+i\beta rq^2,   \\
&&ar_t=-ir_{xx}+\alpha(rq^2)_x-i\beta r^2q,
\end{eqnarray}
where $\alpha, \beta \in \mathbb{R}$, and $a$ is any constant. We have the following discrete symmetry transformations.
For the real case,
\begin{equation}
q(t,x)=k r^{\prime}( \epsilon_{2} t, \epsilon_{1} x), \quad r(t,x)=k q^{\prime}( \epsilon_{2} t, \epsilon_{1} x),
\end{equation}
where $\epsilon_1=\epsilon_2=-1$ and $k^2=1$.  When $q^{\prime}=q$ and $r^{\prime}=r$, these discrete transformations produce the nonlocal reduction $r(t,x)=k q(-t,-x)$ \cite{AbMu3}.

\noindent For the complex case,
\begin{equation}
q(t,x)=k \bar{r}^{\prime}( \epsilon_{2} t, \epsilon_{1} x), \quad r(t,x)=k \bar{q}^{\prime}( \epsilon_{2} t, \epsilon_{1} x),
\end{equation}
where $\epsilon_1=1$, $\bar{a}\epsilon_2=a$, and $k^2=1$. From these discrete transformations we have different nonlocal reductions when $q^{\prime}=q$ and $r^{\prime}=r$ with different values of $\epsilon_{1}=\pm 1$ and $\epsilon_{2}=\pm 1$; $\displaystyle r(t,x)=k\, \bar{q}( -t, x)$ with $\bar{a}=a$,  $\displaystyle r(t,x)=k\, \bar{q}( t, -x)$ with $\bar{a}=-a$, and $\displaystyle r(t,x)=k\, \bar{q}( -t, -x)$ with $\bar{a}=-a$.

\vspace{0.3cm}

\noindent \textbf{(f)}\, \textbf{Loop-soliton System}: This system \cite{AbCl}, \cite{Konno} is given by
\begin{eqnarray}
&&aq_t+\frac{\partial^2}{\partial x^2}\Big[\frac{q_x}{(1-rq)^{3/2}}\Big]=0,    \\
&&ar_t+\frac{\partial^2}{\partial x^2}\Big[\frac{r_x}{(1-rq)^{3/2}}\Big]=0.
\end{eqnarray}
\noindent We have the following discrete symmetry transformations.

\noindent For the real case,
\begin{equation}
q(t,x)=k r^{\prime}( \epsilon_{2} t, \epsilon_{1} x), \quad r(t,x)=k q^{\prime}( \epsilon_{2} t, \epsilon_{1} x),
\end{equation}
where $\epsilon_1=\epsilon_2=-1$ and $k^2=1$. When $q^{\prime}=q$ and $r^{\prime}=r$, these discrete transformations produce the nonlocal reduction $r(t,x)=k q(-t,-x)$ \cite{AbMu3}.

\noindent For the complex case,
\begin{equation}
q(t,x)=k \bar{r}^{\prime}( \epsilon_{2} t, \epsilon_{1} x), \quad r(t,x)=k \bar{q}^{\prime}( \epsilon_{2} t, \epsilon_{1} x),
\end{equation}
where $\bar{a}\epsilon_1\epsilon_2=a$, and $k^2=1$. These discrete transformations produce different nonlocal reductions when $q^{\prime}=q$ and $r^{\prime}=r$ with different values of $\epsilon_{1}=\pm 1$ and $\epsilon_{2}=\pm 1$; $\displaystyle r(t,x)=k\, \bar{q}( -t, x)$ with $\bar{a}=-a$,  $\displaystyle r(t,x)=k\, \bar{q}( t, -x)$ with $\bar{a}=-a$, and $\displaystyle r(t,x)=k\, \bar{q}( -t, -x)$ with $\bar{a}=a$.

\vspace{0.3cm}

\textbf{(g)}\, \textbf{Hydrodynamic type of systems: Shallow water waves}

Recently we studied the reductions in equations of hydrodynamic type \cite{hydro} and obtained several examples of nonlocal version of these equations. An example of equations of hydrodynamic type is  the shallow water waves system \cite{GurZhel}
\begin{eqnarray}\label{sww}
&&a q_{t}=(q+r)q_x+q\,r_x,\\
&&a r_{t}=(q+r)r_x+r\,q_x.
\end{eqnarray}
Here $a$ is a nonzero constant. The discrete transformations which leave this system invariant are following. For the real case,
\begin{eqnarray}
r(t,x)=k\, q^{\prime}(\epsilon_{2} t, \epsilon_{1} x),\\
q(t,x)=k\, r^{\prime}(\epsilon_{2} t, \epsilon_{1} x),
\end{eqnarray}
where $k=\epsilon_{1} \epsilon_{2}$. For the complex case
\begin{eqnarray}
r(t,x)=k\, \bar{q}^{\prime}(\epsilon_{2} t, \epsilon_{1} x),\\
q(t,x)=k, \bar{r}^{\prime}(\epsilon_{2} t, \epsilon_{1} x),
\end{eqnarray}
where $\bar{a}\, k\,\epsilon_{1} \epsilon_{2}=a$. In both cases $k^2=\epsilon_{1}^2=\epsilon_{2}^2=1$ \cite{hydro}.

\vspace{0.3cm}
\noindent
 If we let $q^{\prime}=q$ and $r^{\prime}=r$ we get  the special discrete symmetry transformations which lead to the local and nonlocal reductions. When $q$ and $r$ are real variables we have $r(t,x)=k\, q(\epsilon_{2} t, \epsilon_{1} x)$ then the reduced equation is
\begin{equation}
a q_{t}(t,x)=(q(t,x)+k q(\epsilon_{2} t, \epsilon_{1} x)) q_{x}(t,x)+k q(t,x)\,q_{x}(\epsilon_{2} t, \epsilon_{1} x),
\end{equation}
provided that $k=\epsilon_{1} \epsilon_{2}$ and $a$ is real.

\vspace{0.3cm}
\noindent
When $q$ and $r$ are complex variables we have $r(t,x)=k\, \bar{q}(\epsilon_{2} t. \epsilon_{1} x)$ then the reduced equation is
\begin{equation}
a q_{t}(t,x)=(q(t,x)+k \bar{q}(\epsilon_{2} t, \epsilon_{1} x)) q_{x}(t,x)+k q(t,x)\,\bar{q}_{x}(\epsilon_{2} t, \epsilon_{1} x),
\end{equation}
provided that $\bar{a} k\,\epsilon_{1} \epsilon_{2}=a$ \cite{hydro}.

\textbf{(h)}\, \textbf{Fordy-Kulish Equations}

Let $q^{\alpha}(t,x)$ and $r^{\alpha}(t,x)$ be the complex dynamical variables where $\alpha=1,2,\cdots,N$, then the Fordy-Kulish (FK) integrable system
is given by \cite{fk}

\begin{eqnarray}
a q^{\alpha}_{t}&=& q^{\alpha}_{xx}+ R^{\alpha}\,_{\beta \gamma -\delta}\, q^{\beta}\,q^{\gamma}\, r^{\delta}, \label{denk51}\\
-a r^{\alpha}_{t}&=& r^{\alpha}_{xx}+ R^{-\alpha}\,_{-\beta \gamma \delta}\, r^{\beta}\,r^{\gamma}\, q^{\delta}, \label{denk61}
\end{eqnarray}
where $R^{\alpha}\,_{\beta \gamma -\delta},R^{-\alpha}\,_{-\beta -\gamma \delta}$ are the curvature tensors of a Hermitian symmetric space with
\begin{equation}
(R^{\alpha}\,_{\beta \gamma -\delta})^{\star}=R^{-\alpha}\,_{-\beta -\gamma \delta},\label{prop}
\end{equation}
and $a$ is a complex number. Here we use the summation convention, i.e., the repeated indices are summed up from 1 to $N$. These equations are known as the FK system which is integrable in the sense that they are obtained from the zero curvature condition of a connection defined on a Hermitian symmetric space. The FK equations (\ref{denk51}) and (\ref{denk61}) are invariant under the discrete transformations
\begin{eqnarray}
&&r^{\alpha}(t,x)=k\, \bar{q}^{\prime \alpha}(\epsilon_{1} t, \epsilon_{2} x), \\
&&q^{\alpha}(t,x)=k \,\bar{r}^{\prime \alpha}(\epsilon_{1} t, \epsilon_{2} x),
\end{eqnarray}
where $k^2=\epsilon_{1}^2=\epsilon_{2}^2=1$ and $\bar{a} \epsilon_{2}=-a$. If we let $r^{\prime \alpha}=r^{\alpha}$ and $q^{\prime \alpha}=q^{\alpha}$ we obtain the special discrete symmetry transformations and hence  the nonlocal reductions
$r^{\alpha}(t,x)=k\, \bar{q}^{\alpha}(\epsilon_{1} t, \epsilon_{2} x)$  \cite{GursesFK}. Then the reduced nonlocal FK equations are

\begin{equation}
a q^{\alpha}_{t}(t,x)=q^{\alpha}_{xx}(t,x)+ k\,R^{\alpha}\,_{\beta \gamma -\delta}\, q^{\beta}(t,x)\,q^{\gamma}(t,x)\, \bar{q}^{\delta}(\epsilon_{1} t, \epsilon_{2} x).
\end{equation}

\section{Conclusion}

In this work we showed that the discrete symmetries of systems of integrable equations are important in finding the nonlocal reductions. For this reason we started first with the scale symmetry transformations of real and complex dynamical systems.  Discrete symmetry transformations are special cases of the scale transformations. There are two different types of discrete symmetry transformations both for real and complex dynamical variables. Using this fact we can find all discrete symmetry transformations of the system of equations.
Among these discrete symmetry transformations the first types are the origins of the nonlocal reductions of these systems. We showed that a special discrete symmetry transformation of the first type produces all the well known nonlocal reductions.

\section{Acknowledgment}
 This work is partially supported by the Scientific
and Technological Research Council of Turkey (T\"{U}B\.{I}TAK).

\end{document}